\input{aipcheck.tex}

\newcommand\sps{\space\space\space\space}
\documentclass[final
  ]
  {aipproc}

\usepackage{axodraw}

\layoutstyle{6x9}

\SetInternalRegister\hbadness{8000}

\newcommand\doingARLO[2][]{%
  \ifx\mmref\undefined #1\else #2\fi
}

\begin{document}

\title
      [$Q$-genesis]
      {Baryogenesis by Heavy Quarks: $Q$-genesis}


\author{Jihn E. Kim}{
  address={Department of Physics and Astronomy, Seoul National
  University, Seoul 151-747, Korea},
  email={jekim@phyp.snu.ac.kr},
  thanks={Talk presented at CICHEP-II, Cairo, Egypt, Jan. 15, 2006.}
}


\copyrightyear  {2001}

\begin{abstract}
In this talk, I present a new mechanism for baryogenesis the
Q-genesis in which  the heavy quarks are the source of baryon number
\cite{Qgenesis}. There exists a narrow allowed region for the
Q-genesis.
\thanks{Talk presented at CICHEP-II, Cairo, Egypt, Jan. 15, 2006.
}
\end{abstract}

\date{\today}

\maketitle

\section{Introduction}

From the observed facts in the heaven, astrophysics and cosmology
deal with cosmic microwave background radiation (CMBR), abundant
light elements, galaxies and intergalactic molecules, and dark
matter (DM) and dark energy (DE) in the universe. In this talk, I
present a recent work \cite{Qgenesis} regarding the light elements
in this list whose source can be baryon number ($B$) from heavy
quark decay.

Sakharov's three conditions for generating $\Delta B\ne 0$ from a
baryon symmetric universe are
\begin{itemize}
\item Existence of $\Delta B\ne 0$ interaction,
\item C and CP violation, and
\item Evolution in a nonequilibrium state.
\end{itemize}
GUTs seemed to provide the basic theoretical framework for
baryogenesis, because in most GUTs $\Delta B\ne 0$  interaction is
present. Introduction of C and CP violation is always possible if
not forbidden by some symmetry. The third condition on the
non-equilibrium state evolution can be possible in the evolving
universe but it has to be checked with specific interactions.

Thus, a cosmological evolution with  $\Delta B\ne 0$  and C and CP
violating particle physics model can produce a nonvanishing $\Delta
B$. The problem is $\lq\lq$How big is the generated $\Delta B$".
Here, for nucleosynthesis, we need
\begin{equation}
 \Delta B\simeq 0.6\times 10^9 n_\gamma.
\end{equation}
For example, the SU(5) GUT with X and Y gauge boson interactions are
not generating the needed magnitude when applied in the evolving
universe. In the SU(5) GUT, two quintet Higgs are needed for the
required magnitude. With this scenario, GUTs with colored scalars
seemed to be the theory for baryogenesis for some time.

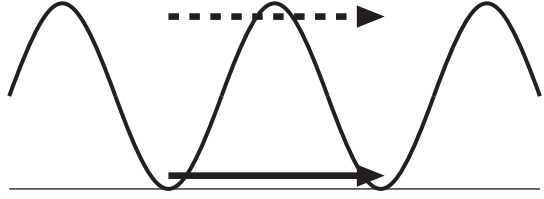
\begin{figure}[t]
\begin{picture}(400,80)(0,0)

\Line(90,5)(290,5)
 \SetWidth{1.5}\Photon(90,40)(290,40){35}{2.5}
 \SetWidth{2.5} \LongArrow(150,10)(230,10)
 \DashLine(150,70)(225,70){4} \LongArrow(225,70)(230,70)

\end{picture}
\caption{Vacuum tunneling at high (dashed arrow) and low (solid
arrow) temperatures.}\label{VacTun}
\end{figure}

\begin{figure}[h]
\begin{picture}(400,150)(0,0)
\CArc(200,70)(10,0,360) \SetWidth{1}
 \LongArrow(135,70)(188,70) \Text(130,70)[r]{$q_2^r$}
 \LongArrow(265,70)(212,70) \Text(270,70)[l]{$l_3$}
 \LongArrow(200,135)(200,82) \Text(200,143)[c]{$q_1^r$}
 \LongArrow(200,5)(200,58) \Text(200,-3)[c]{$q_3^r$}

 \LongArrow(260,105)(211,75) \Text(265,108)[l]{$l_2$}
 \LongArrow(140,105)(189,75) \Text(135,108)[r]{$q_1^y$}
 \LongArrow(260,35)(211,65) \Text(265,42)[l]{$q_3^y$}
 \LongArrow(140,35)(189,65) \Text(135,42)[r]{$q_2^g$}

 \LongArrow(235,130)(206,79) \Text(237,132)[l]{$l_1$}
 \LongArrow(165,130)(194,79) \Text(163,132)[r]{$q_1^g$}
 \LongArrow(235,10)(206,61) \Text(237,12)[l]{$q_3^g$}
 \LongArrow(165,10)(194,61) \Text(163,12)[r]{$q_2^y$}
\end{picture}
\caption{The sphaleron process via 't Hooft
interaction.}\label{Sphaleron}
\end{figure}
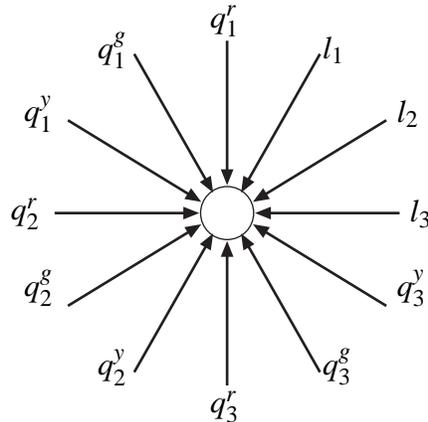

But high temperature QFT aspects changed this view completely. The
spontaneously broken electroweak sector of the standard model (SM)
does not allow instanton solutions. When the SU(2)$_W$ is not
broken, there are electroweak SU(2) instanton solutions. Tunneling
via these electroweak instantons is extremely suppressed,
$\sim\exp(-2\pi/\alpha_w)$. This tunneling amplitude is the zero
temperature estimate. At high temperature where the electroweak
phase transition occurs, the transition rate can be huge, and in
cosmology this effect must be considered \cite{Kuzmin}. The
tunneling amplitude due to sphaleron effect is large at high and
small at low temperatures as shown in Fig. \ref{VacTun}.
This sphaleron effect transforms SU(2) doublets. The 't Hooft vertex
for this process must be a SM singlet, which is shown in Fig.
\ref{Sphaleron}.

The baryon number violating interaction washes out the baryon
asymmetry produced during the GUT era. Since the above sphaleron
interaction violates $B+L$ but conserves      $B-L$, if there were a
net $B-L$, there results a baryon asymmetry below the weak scale.
The partition of $(B-L)$ into $B$ and $L$ below the electroweak
scale is the following if the complete washout of $B+L$ is achieved,
\begin{equation}
\Delta B=\textstyle\frac12 (B-L)_{\rm orig.},\quad \Delta
L=\textstyle\frac12 (B-L)_{\rm orig.}
\end{equation}
$B-L$ must be present in the beginning. If so, the leptogenesis uses
this transformation of the $(B-L)$ number(obtained from heavy $N$
decays) to baryon number. The  $\nu$-genesis also uses this
transformation.

Thus, in some $L$ to $B$ transformation models, we need to generate
a net $(B-L)$ number at the GUT scale ($Q$-genesis does not need
this). The SU(5) GUT conserves $B-L$ and hence cannot generate a net
$B-L$.

So, for the baryon number generation one has to go beyond the SU(5)
GUT. There exist three examples of baryon number generation  from
fermion sources:
 (1) Leptogenesis \cite{leptogenesis}, (2) Neutrino genesis \cite{nugenesis},
  and (3) Q-genesis \cite{Qgenesis}.

\section{Introduction of heavy quarks}

We note that SU(2) singlets avoid the sphaleron process and hence
singlet quarks survive the electroweak era. With this idea, we must
pursue along the line of heavy quarks that achieves
\begin{itemize}
\item Heavy quarks must mix with light quarks so that after the electroweak
phase transition they can generate the quark number ($B/3$).
\item
They must be sufficiently long lived.
\item Of course, a correct
order of $\Delta B \ne 0$ should be generated.
\end{itemize}
The SU(2) singlet quarks were considered before in connection with
(i) flavor changing neutral currents (FCNC) \cite{KK}, and recently
for the BELLE data \cite{Moro}.

For the absence of FCNC at tree level, the electroweak isospin $T_3$
eigenvalues must be the same. Thus, introducing L-hand quark
singlets will potentially introduce the FCNC problem. But in most
discussions, the smallness of mixing angles with singlet quarks has
been overlooked. Since the quark singlets can be superheavy compared
to 100 GeV, the small mixing angles are natural, rather than being
unnatural.

For definiteness, let us consider $Q_{\rm em}= -\frac13$ heavy quark
$D$ for which we must satisfy
\begin{enumerate}
\item $\Delta D$ generation mechanism is possible,
\item  $10^{-10} {\rm s} < \tau_D < 1$ s,
\item  Sphaleron should not wash out all $\Delta D$, and
\item  The FCNC bound is satisfied.
\end{enumerate}
Theoretically, is it natural to introduce such a heavy quark(s)? It
is so. For example,
   in E6 GUT there exist $D$s in ${\bf 27}_F$.  Trinification GUT
   also has particle $D$, viz.
\begin{eqnarray}
  &&{\bf  27 \to 16 + 10 + 1 \to 10 + 5^* + 1 + 5 + 5^* + 1}\nonumber\\
 &&\to (q+u^c+e^c) + N_5 + (l+d^c) + (D+L_2)  + (D^c+L_1) +N_{10}
\end{eqnarray}
When we consider this kind of vector-like quarks $(D+D^c)$, there
are three immediate related physical problems to deal with: heavy
quark axion, the Nelson-Barr type, and FCNC. But here, we will
consider the FCNC problem only.

We will consider one family first for $d$-type, $(t,\ b)^T_L,\ b_R$.
It gives all the needed features. The mass matrix is
\begin{equation}
M_{-1/3}=\left(
\begin{array}{cc}
m&J\\ 0&M
\end{array}\right)
\end{equation}
The entry 0 in the above is a natural choice, since it can be
achieved by redefinition of R-handed singlets. The above mass matrix
can be diagonalized by considering $MM^\dagger$, to give
\begin{equation}
\left(\begin{array}{c} |m_b|^2\\ |m_D|^2 \end{array}\right)=
\textstyle\frac12(|M|^2+|m|^2+|J|^2)\mp\sqrt{[(|M|+|m|)^2+|J|^2]
\cdot[(|M|-|m|)^2+|J|^2]}
\end{equation}
which tends to $m_b\to m$ and $m_D\to M$ in the limit of $M^2\gg
|m|^2,|mJ|$. So, with vanishing phases, we have the following
eigenstates,
\begin{equation}
|b\rangle\simeq\left(
\begin{array}{c}
1\\ -J/M
\end{array}\right),\quad |D\rangle\simeq\left(
\begin{array}{c}
J/M\\ 1
\end{array}\right).
\end{equation}
Since $J$ is the doublet VEV and $M$ is a parameter or a singlet
VEV, the mixing angle can be sufficiently small. This is the
well-known decoupling of vectorlike quarks. It can be generalized to
three ordinary quarks and $n$ heavy quarks. The $(3+n)\times(3+n)$
matrix
$$
M_{-1/3}=\left(
\begin{array}{cc}
M_d&J\\ J^\prime&M_D
\end{array}\right)
$$
can take the following form by redefining R-handed $b$ and $D$
fields,
\begin{equation}
M_{-1/3}=\left(
\begin{array}{cc}
M_d&J\\ 0&M_D
\end{array}\right)
\end{equation}
For an easy estimate, below we express $J$ as
$$
J=fm_b.
$$

\section{Q-genesis by heavy quarks}

\begin{figure}[t]
\begin{picture}(400,140)(0,0)
\SetWidth{0.9} \ArrowLine(50,70)(100,70)  \Text(45,70)[r]{$X_1$}
 \ArrowLine(170,130)(100,70)\Text(170,125)[l]{$u^c$}
 \ArrowLine(170,10)(100,70)\Text(170,20)[l]{$D^c$}

\ArrowLine(210,70)(260,70)  \Text(205,70)[r]{$X_1$}
\ArrowLine(260,70)(295,100)\ArrowLine(330,130)(295,100)
 \Text(325,130)[r]{$u^c$}
\ArrowLine(260,70)(295,40)\ArrowLine(330,10)(295,40)
 \Text(325,10)[r]{$D^c$}

\DashLine(295,100)(295,40){4} \Text(300,70)[l]{$X_1,\ X_2$}
\Text(280,90)[r]{$e^c$} \Text(280,48)[r]{$u^c$}

  \end{picture}
\caption{Diagrams contributing to $X\to
u^c+D^c$.}\label{fig:Interfere}
\end{figure}
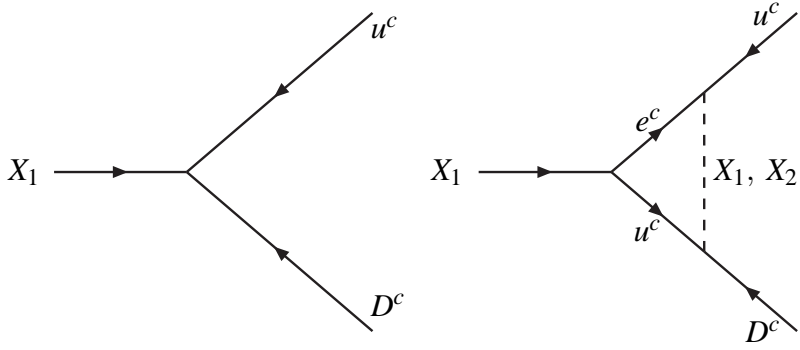

Here, we generate the D number as usual in GUTs cosmology through
the Sakharov mechanism, and in the end we will identify the $D$
number as the $3B$ number. The relevant interaction we introduce is
\begin{equation}
g_{Di}X_iu^c D^c+ g_{ei}X_i^*u^ce^c+{\rm h.c.}
\end{equation}
which can introduce $\Delta D$ through the interference term from
Fig. \ref{fig:Interfere} plus self energy diagrams.

The interference is needed to generate a nonvanishing $D$ number.
The cross term contributes as $g_{e1}^*g_{e2}g_{D1}^* g_{D2}$. If we
allow arbitrary phases in the Yukawa couplings, the relative phases
of $g_{D1}$ and $g_{D2}$ can be cancelled only by the relative phase
redefinition of $X_2$ and $X_1$. The same applies to $g_{e1}$ and
$g_{e2}$. Thus, the phase $\eta$ appearing in
$g_{e1}^*g_{e2}g_{D1}^* g_{D2}$ is physical. The $D$ number
generated in this way is
\begin{eqnarray}
&&\frac{n_D}{n_\gamma}\simeq 0.5\times 10^{-2}\epsilon\\
&&\epsilon \sim 10^{-2}\frac{\eta}{8\pi}[F(x)-F(1/x)],\quad
x=\frac{M_{X_1}}{M_{X_2}}
\end{eqnarray}
where $F(x)=1-x\ln(1+(1/x))$.

\section{Lifetime estimate}

The dominant decay of $D$ proceeds via
\begin{equation}
D \to tW, bZ, bH^0
\end{equation}
from which we obtain
\begin{equation}
\Gamma_D=\frac{\sqrt2 G_F}{8\pi}|J|^2m_D,\quad
\epsilon=\frac{fm_b}{m_D}
\end{equation}
The lifetime of $D$ must be made longer than $2x10^{-11}$ s
(beginning of electroweak phase transition), and it should be made
shorter than 1 s (beginning of nucleosynthesis), i.e. $ 2\times
10^{-11}\ {\rm s} <\tau_D< 1\ {\rm s}$, which gives
\begin{equation}
  \frac{1}{(10^6m_{D,GeV})^{3/2}}<|\epsilon|<\frac{1}{
  (2.7\times 10^2m_{D,GeV})^{3/2}},\quad m_{D,GeV}=\frac{m_D}{\rm
  GeV}.
\end{equation}
The mixing of $D$ with $b$ is of order $\epsilon$. For one period of
oscillation, we expect that a fraction $|\epsilon|^2$ of $D$ is
expected to transform to $b$. Since the period keeping the
electroweak phase in cosmology is of order,
$$
\frac{1}{H}=\sqrt{\frac{3M_P^2}{\rho}}\simeq\sqrt{\frac{3}{g_*}}
\frac{M_W}{M_P^2}.
$$
Thus, the following fraction is expected to be washed out via $D$
oscillation into $b$,
$$
\sim \frac{M_Pm_b^2}{M_W^2m_D}f^2\simeq
10^{16}\frac{f^2}{m_{D,GeV}}.
$$
For $m_D$ of order $> 10^6$ GeV, we need $f<10^{-5}$. In this range,
some heavy quarks are left and the sphaleron does not erase the
remaining $D$ number.

\section{FCNC}

For the FCNC, we consider the following processes which are compared
with the existing experimental bounds
\begin{eqnarray}
&&Z\to b\bar b,\quad |z_{bb}|=0.996\pm 0.005\quad \to\quad
|\epsilon|\le 0.009\nonumber\\
 && B\to X_sl^+l^-,\quad {\rm from\ tree}\quad|z_{sb}|=\frac{J_bJ_s}{m_D^2}<
 1.4\times 10^{-3}\nonumber\\
&& K^+\to\pi^+\nu\bar\nu,\quad |z_{sd}|\le 7.3\times 10^{-6}
\end{eqnarray}
where the last bound comes from
$$
\frac{{\rm Br.}(K^+\to \pi^+\nu\bar\nu)_{FCNC}}{{\rm
Br.}(K^+\to\pi^0 e^+\nu)}=\frac{3|z_{sd}|^2}{2\lambda^2}\le 2\times
10^{-9}.
$$
Therefore, from the FCNC processes, we obtain the bound on the
coupling $f$,
\begin{equation}
 \frac{1}{(4.8\times 10^9\sqrt{m_{D,GeV}})}<|f|<\frac{1}{
  (2.1\times 10^4\sqrt{m_{D,GeV}})}.
\end{equation}

\begin{figure}[t]
\begin{picture}(400,250)(0,-10)
\SetWidth{1} \Line(50,20)(330,20) \Line(50,250)(330,250)
\Line(50,20)(50,250) \Line(330,20)(330,250) \SetWidth{0.5}
\Line(120,20)(120,25)\Line(190,20)(190,25) \Line(260,20)(260,25)
\Line(120,250)(120,245) \Line(190,250)(190,245)
\Line(260,250)(260,245) \Text(50,10)[c]{$0.1$} \Text(120,10)[c]{$1$}
\Text(190,10)[c]{$10$} \Text(260,10)[c]{$10^2$}
\Text(330,10)[c]{$10^3$}

\Line(50,58.3)(55,58.3)\Line(50,96.7)(55,96.7)\Line(50,135)(55,135)
\Line(50,173.3)(55,173.3)\Line(50,211.7)(55,211.7)
\Line(330,58.3)(325,58.3)\Line(330,96.7)(325,96.7)
\Line(330,135)(325,135)
\Line(330,173.3)(325,173.3)\Line(330,211.7)(325,211.7)
\Text(45,20)[r]{$10^{-18}$} \Text(45,58.3)[r]{$10^{-16}$}
\Text(45,96.7)[r]{$10^{-14}$} \Text(45,135)[r]{$10^{-12}$}
\Text(45,173.3)[r]{$10^{-10}$} \Text(45,211.7)[r]{$10^{-8}$}
\Text(45,250)[r]{$10^{-6}$}

\LongArrow(300,195)(290,185)\LongArrow(200,153)(210,169)

\Text(10,150)[r]{$\epsilon$}
 \Text(190,-5)[c]{ $m_D$(TeV)}

\Line(90,237.3)(210,237.3)\Text(200,230)[l]{FCNC} \SetWidth{1.2}
\Line(50,135)(330,20) \DashLine(50,237.9)(330,124){5}
\DashLine(50,219.8)(330,178.3){2} \SetWidth{1.6}
\LongArrow(110,237.3)(110,225)
\LongArrow(130,237.3)(130,225)\LongArrow(150,237.3)(150,225)
\LongArrow(170,237.3)(170,225)\LongArrow(190,237.3)(190,225)

\LongArrow(60,219)(60,209)\LongArrow(70,217)(70,207)
\LongArrow(80,215)(80,205) \LongArrow(140,201)(140,191)
\LongArrow(150,197)(150,187) \LongArrow(160,192)(160,182)

\LongArrow(90,119)(90,129)\LongArrow(100,114.5)(100,124.5)
\LongArrow(110,110)(110,120)

\Text(220,200)[l]{Mixing to sphaleron} \Text(250,150)[r]{EW phase
transition} \Text(130,110)[l]{BBN}

\end{picture}
\caption{Allowed parameter space in terms of mixing angle and $D$
quark mass.}\label{Bounds}
\end{figure}
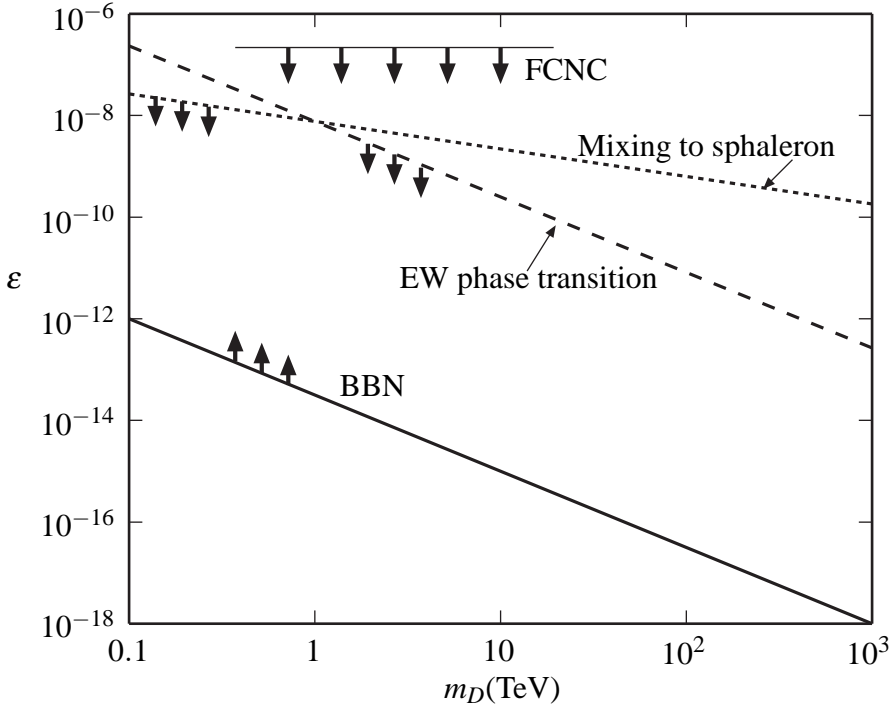

Fig. \ref{Bounds} summarizes the allowed parameter space in terms of
mixing angle and $D$ quark mass. The remarkable fact is that there
exist a band allowed by the current data.

\section{$Z_2$ symmetry}

To implement a small $f$ naturally, we can impose a discrete
symmetry such as a parity symmetry $Z_2$,
\begin{eqnarray}
Z_2\ :\ \left\{\begin{array}{l} b_{L,R}\to b_{L,R},\  D_{L,R}\to D_{L,R}\\
\phi\to\phi,\  \varphi\to-\varphi
\end{array}
\right.
\end{eqnarray}
where $\phi$ and $\varphi$ are Higgs doublets giving mass and
mixing, respectively. The discrete symmetry forbids a mixing between
doublets. We introduce a soft term, violating the discrete symmetry,
which can mix them. Consistently with the discrete symmetry except
the soft term, the potential is given by
\begin{eqnarray}
V &=& (m_\delta^2\varphi^*\phi+{\rm h.c.}) -\mu^2\phi^*\phi
+M_\varphi^2\varphi^*\varphi\nonumber\\
   && +\lambda_1(\varphi^*\varphi)^2+\lambda_2(\phi^*\phi)^2+\cdots
\end{eqnarray}
The soft term violates the $Z_2$ symmetry. If the $Z_2$ is exact,
there is no mixing between $D$ and $b$ and then $D$ is absolutely
stable: $\langle\phi\rangle = v\ne 0$, and $\langle\varphi\rangle =
 0$. But the existence of the soft term violates the $Z_2$ and a
tiny VEV,  $\langle\varphi\rangle $, is generated. The estimate of
mixing is
$$
\langle\varphi\rangle =\frac{m_\delta^2 v}{M_\varphi^2},\ f_{\rm
off}q_L\varphi D_R\quad\to\quad J=f_{\rm off}\frac{ m_\delta^2
v}{M_\varphi^2}
$$
from which we estimate
\begin{equation}
f=\frac{\sqrt2 f_{\rm off} m_\delta^2}{f_b M_\varphi^2}
\end{equation}
which can lead to naturally small values of $|J|=fm_b$ and
$\epsilon=J/M_D$. Typical value for $m_D$ is given by constraints.
The mass  $M_\varphi^2$ can be superheavy.

Another point to be stressed, but unrelated to $Q$-genesis, is that
we can construct a DM model of $D$ with an exact $Z_2$ symmetry.

A related idea in the effective theory, but not the same at the
renormalizable level, can be found in \cite{Hempl}, where $R$-parity
violating SUSY is used with $B$ carrying singlet scalars $S$,
$$
  W_{NR} \sim\frac{1}{M_T} u^cd^cd^c S +{\rm h.c.}
$$
where $T$ is our heavy quark $D$. But here $T$ is much heavier than
$S$. It has the similar idea of evading the electroweak phase
transition era.

\section{Conclusion}

We presented a new mechanism for baryogenesis: the   $Q$-genesis. We
obtained the constraints on the parameter space. With a $Z_2$
discrete symmetry, the smallness of the parameter is naturally
implemented.
\begin{table}
\begin{tabular}{lccl}
\hline
  & $B-L=0$
  & $B-L\ne 0$
  & Sphaleron
    \\
\hline
leptogenesis \cite{leptogenesis} &  & Yes & $(\nu,e)_L$ converts to $B$  \\
$\nu$-genesis \cite{nugenesis}& Possible & Possible &  $(\nu,e)_L$ converts to $B$  \\
$Q$-genesis \cite{Qgenesis}& Possible & Possible & $Q$ decay produces $B$ \\
\hline
\end{tabular}
\caption{Comparison of a few baryogenesis mechanisms.}
\label{tb:compare}
\end{table}

Finally,  in Table \ref{tb:compare}, this $Q$-gensis mechanism is
compared with other baryogenesis mechanisms. Leptogenesis and
$\nu$-genesis seem to be the plausible ones since the observed
neutrino masses need R-handed neutrinos. Survival hypothesis sides
with leptogenesis. $Q$-genesis depends on the unobserved $Q$, but
this may be needed in solutions of the strong CP problem. Also, it
appears in E$_6$ and trinification GUTs. One should consider all
kinds  of SM singlet particles and vector-like representations for
the baryon asymmetry in the universe:\footnote{Note colored scalars
can appear in the Affleck-Dine mechanism \cite{ADine}.}
$$
\begin{array}{lcl}
    N,\ \nu_R   &:& {\rm fermion\ singlets}\\
        Q_L+Q_R&:& {\rm vectorlike\ fermion}\\
        N_L+N_R&:& {\rm vectorlike\ fermion\ with\ Dirac\ mass}\\
         S&:& {\rm singlet\ scalar\ carrying\ \it B\rm\ number}
\end{array}
$$
In Table \ref{tb:compare}, the $B-L$ number is that of light
fermions. So, in the neutrino-genesis, one counts the R-handed light
neutrino($\nu_R$) number also in the $B-L$ number. In the
$Q$-genesis, whatever sphaleron does on the light lepton number,
still there exists baryon number generation from the $Q$ decay,
which occurs independently from the light fermion number.

\begin{theacknowledgments}
This work is supported in part by  the KRF grants, No.
R14-2003-012-01001-0, No. R02-2004-000-10149-0, and No.
KRF-2005-084-C00001.
\end{theacknowledgments}


\doingARLO[\bibliographystyle{aipproc}]
          {\ifthenelse{\equal{\AIPcitestyleselect}{num}}
             {\bibliographystyle{arlonum}}
             {\bibliographystyle{arlobib}}
          }
\bibliography{sample}

\begin{thebibliography}{99}

\def\apj#1#2#3{Astrophys.\ J.\ {\bf #1} (#3) #2}
\def\ijmp#1#2#3{Int.\ J.\ Mod.\ Phys.\ {\bf #1} (#3) #2}
\def\mpl#1#2#3{Mod.\ Phys.\ Lett.\ {\bf A#1} (#3) #2}
\def\npb#1#2#3{Nucl.\ Phys.\ {\bf B#1} (#3) #2}
\def\plb#1#2#3{Phys.\ Lett.\ {\bf B#1} (#3) #2}
\def\prd#1#2#3{Phys.\ Rev.\ {\bf D#1} (#3) #2}
\def\prl#1#2#3{Phys.\ Rev.\ Lett.\ {\bf #1} (#3) #2}
\def\prt#1#2#3{Phys.\ Rep.\ {\bf #1} (#3) #2}
\def\sjnp#1#2#3{Sov.\ J.\ Nucl.\ Phys.\ {\bf #1} (#3) #2}
\def\zp#1#2#3{Z.\ Phys.\ {\bf #1} (#3) #2}
\def\jhep#1#2#3{JHEP\ {\bf #1} (#3) #2}
\def\ephjc#1#2#3{Europhys. J. C\ {\bf #1} (#3) #2}

\bibitem{Qgenesis} H. D. Kim, J. E. Kim, and T. Morozumi,
\plb{616}{108}{2005}.

\bibitem{Kuzmin} V. A. Kuzmin, V. A. Rubakov, and M. E. Shaposhnikov,
\plb{155}{36}{1985}.

\bibitem{leptogenesis} M. Fukugita and T. Yanagida, \plb{174}{45}{1986}.

\bibitem{nugenesis} K.~Dick,
M.~Lindner, M.~Ratz and D.~Wright, Phys.\ Rev.\ Lett.\  {\bf 84},
4039 (2000).

\bibitem{KK} K. Kang and J. E. Kim, \plb{64}{93}{1976}.

\bibitem{Moro} L. T. Handoko and T. Morozumi, \mpl{10}{309}{1995} and
\mpl{10}{1733}{1995}(E).

\bibitem{Hempl}  S. Davidson and R. Hempfling, \plb{391}{287}{1997}.

\bibitem{ADine} I. Affleck and M. Dine, \npb{249}{361}{1985}.

\end{thebibliography}

\end{document}